%% file: paper_scuba_graph.tex
\documentclass[a4paper,oneside,11pt]{article}

\input{preamble/packages.tex}

\input{preamble/commands.tex}

\input{preamble/abbreviations.tex}

\graphicspath{
    {figures/}
}

\title{Scuba Diving Graphs}

\author[1]{Alexander M. Esser\,\orcidlink{0000-0002-5974-2637}}

\affil[1]{University of Koblenz, Germany}

\affil[ ]{\texttt{alexanderesser@uni-koblenz.de}}

\date{}

\begin{document}
\maketitle

\begin{abstract}
Scuba divers form small and structured social groups that are well suited for analysis using methods from Computational Social Science (CSS). This paper proposes a conceptual graph-based framework for modeling scuba dives as social networks. Divers are represented as vertices and their interactions as edges within a temporal network. The model focuses on three dimensions of interaction: physical distance, communicative distance, and emergency distance, reflecting spatial positioning, effectiveness of underwater communication, and the ability to assist in critical situations. While the present work is conceptual and primarily descriptive, the proposed graph representation may provide a foundation for future empirical studies aimed at improving diver interaction, coordination, and safety.
\end{abstract}

\vfill

\input{parts/introduction}
\input{parts/introduction/outline}

\input{parts/background}
\input{parts/background/scuba_diving}
\input{parts/background/graphs}

\input{parts/related_work}

\input{parts/graph_representation}
\input{parts/graph_representation/dimensions}

\input{parts/graph_representation/example}
\input{parts/graph_representation/vertical_limitations}

\input{parts/conclusion}

\vspace{1em plus 0.5em}

\needspace{5\baselineskip}

\section*{Acknowledgments}
{
\footnotesize
During the preparation of this manuscript, the author used generative AI tools (ChatGPT, DeepL) for spelling and grammar checks and wording suggestions.
The author reviewed and edited all generated content and takes full responsibility for the final text.



\medskip

\noindent Special thanks to Christoph Birkholz for reviewing the scuba-related content, in the best sense of the buddy principle.
\medskip

\noindent Scuba diver icon: Freepik (on flaticon.com).

}

\vspace{1em plus 0.5em}

\bibliographystyle{apalike}
\bibliography{references}

\end{document}

%% file: preamble/packages.tex
\usepackage[american]{babel}
\usepackage{csquotes}

\usepackage{subcaption}
\usepackage{calc}

\usepackage{amssymb}
\usepackage{amstext}

\usepackage{pslatex}
\usepackage[bottom]{footmisc}

\usepackage{setspace}
\usepackage{xspace}
\usepackage{xurl}
\usepackage{authblk}

\usepackage{natbib}

\usepackage{xcolor}

\usepackage[acronym, nomain, section]{glossaries}
\setacronymstyle{long-short}

\usepackage{tabularx}
\usepackage{booktabs}
\usepackage{microtype}
\usepackage{needspace}

\usepackage[hidelinks]{hyperref}
\usepackage{cleveref}
\usepackage{orcidlink}

%% file: preamble/commands.tex
\newcommand{\eg}{e.\,g.\@\xspace}
\newcommand{\ie}{i.\,e.\@\xspace}


%% file: preamble/abbreviations.tex
\newacronym{CRF}{CRF}{Conditional Random Field}
\newacronym{CSS}{CSS}{Computational Social Science}
\newacronym{CVET}{CVET}{Continuing Vocational Education and Training}

\newacronym{DAPT}{DAPT}{Domain-adaptive Pre-training}

\newacronym{DNN}{DNN}{Deep Neural Network}

\newacronym{HMM}{HMM}{Hidden Markov Model}

\newacronym{IE}{IE}{Information Extraction}

\newacronym{LLM}{LLM}{Large Language Model}
\newacronym{LSTM}{LSTM}{Long short-term memory}

\newacronym{NAT}{NAT}{Noise-aware Training}
\newacronym{NER}{NER}{Named Entity Recognition}
\newacronym{NLP}{NLP}{Natural Language Processing}

\newacronym{OCR}{OCR}{Optical Character Recognition}

\newacronym{VET}{VET}{Vocational Education and Training}
\newacronym{VLM}{VLM}{Vision Language Model}

%% file: parts/introduction.tex

\section{Introduction}
\label{sec:introduction}

Scuba divers can be modeled as a dynamic social network. This paper explores how methods from \gls{CSS}, in particular graph-based approaches, can be applied to the field of scuba diving.

Computational Social Science uses computational and data-driven methods to analyze social structures and interactions. Social network analysis focuses specifically on the relationships between social actors, providing a powerful framework to study interrelated data. Social structures are commonly represented as graphs: individuals are modeled as nodes, relationships or interactions between them are modeled as edges \citep{Doerpinghaus2024}.

Perhaps we know this from our group of friends: There often is that one guy holding the group together and taking the initiative to make plans. In graph terms, this person would have a high centrality value.
Various authors have examined centrality in graphs and social structures \cite{Brandes2001,Kintali2008,Gago2012,Doerpinghaus2023b,Doerpinghaus2024,Mangroliya2024}.

\pagebreak[3]
\needspace{5\baselineskip}

Graph-based approaches can also be applied to scuba diving. Beyond the author's personal enthusiasm for diving, it is indeed particularly suitable for applying \gls{CSS} methods:

\vspace{0em plus 0.2em}
\begin{enumerate}
    \item Divers usually move in clearly structured formations (see \Cref{sec:background_scuba_diving_positions}). There are well-defined roles and positions within the group. These arrangements can naturally be represented as graphs. Clear roles are not strictly necessary for the application of CSS methods, but they are helpful, for example, in interpreting graph positions and interaction patterns. While social network analysis often aims to \emph{infer} roles from observations, in scuba diving, the roles are already clearly assigned.

    \item Communication underwater is reduced to the essential minimum. Unlike many other social settings in which extremely large amounts of communication data must be filtered and processed, underwater interaction is limited to standardized hand signals. This makes interactions more explicit and easier to model computationally. In computer science and social sciences, there is consensus that small, dense networks often allow for deeper and more detailed analysis than large, \enquote{poor} networks \citep{Doerpinghaus2024}.\pagebreak[1]

    \item Social interactions in scuba diving, such as distance and responsiveness, are safety-critical factors. Therefore, modeling these relations in a graph structure is not only analytically and scientifically interesting but immediately practically relevant.
\end{enumerate}
\vspace{0em plus 0.2em}

In this sense, scuba diving provides a compact and structured social microcosm which is well suited for applying \gls{CSS} methods, especially graph-based models.

One of the fundamental principles of recreational scuba diving is the \emph{buddy principle}. Divers generally dive in pairs rather than alone, with each diver assigned a buddy.
The buddy system serves several purposes, including mutual responsibility and assistance, supporting communication, and providing immediate help in case of problems.

In social network analysis, many publications focus on \emph{pairwise} interactions, for reasons of analytical simplicity~\citep{Cencetti2021, Doerpinghaus2024}. This aligns well with scuba diving, where the most relevant interactions typically occur between two buddies or between a diver and an instructor.

\needspace{4\baselineskip}

Within a graph representation of a dive group, \emph{three fundamental dimensions} can be distinguished:

\vspace{0em plus 0.2em}
\begin{enumerate}
    \item physical distance (relative spatial position),
    \item communicative distance (effectiveness and clarity of communication),
    \item emergency distance (ability to intervene in case of a problem).
\end{enumerate}
\vspace{0em plus 0.2em}

These three dimensions are highly correlated but not always identical. Two divers may be physically close but unable to communicate effectively due to murky water and low visibility.

Such multidimensional data lead to complex network structures, without established best practices for analysis or even description \citep{Lemercier2015}. In the proposed framework, this complexity is addressed by treating the three dimensions as separate but interrelated layers of a multidimensional network. These layers can be compared to detect mismatches between them, such as situations in which divers are physically close but unable to communicate effectively or to provide rapid emergency assistance.

\pagebreak[2]

This paper proposes a conceptual framework for representing scuba dives as multidimensional graphs along these three dimensions. It demonstrates how \gls{CSS} methods can be transferred to an unconventional but structurally well-suited field. The approach is currently conceptual; however, it opens perspectives for future empirical evaluation and practical application.

%% file: parts/introduction/outline.tex

The remainder of this paper is structured as follows:
\Cref{sec:background} introduces the necessary background on scuba diving and graph theory.
\Cref{sec:related_work} reviews related work in \gls{CSS} and graph theory.
\Cref{sec:graph_representation} presents the proposed graph-based representation of scuba dives.
Finally, \Cref{sec:conclusion} concludes the paper and outlines potential directions for future research.

\pagebreak[2]

%% file: parts/background.tex

\section{Background}
\label{sec:background}

This section provides background information on scuba diving and graph theory.

%% file: parts/background/scuba_diving.tex

\subsection{Scuba Diving}
\label{sec:background_scuba_diving}

Scuba diving is typically performed in small groups following well-established safety principles and organizational structures. As mentioned in the introduction, one of the most fundamental rules is the \emph{buddy principle}, according to which divers do not dive alone but always in pairs. Each diver is assigned a buddy who monitors the other diver's condition, assists in communication, and can provide help in case of problems~\citep{PADIBuddyPrinciple, PADIOWDManual}.

\subsubsection{Positions in the group}
\label{sec:background_scuba_diving_positions}

\begin{figure}[tbp]
    \centering

    \begin{subfigure}{1.0\hsize}
        \centering
        \includegraphics[width=0.8\linewidth]{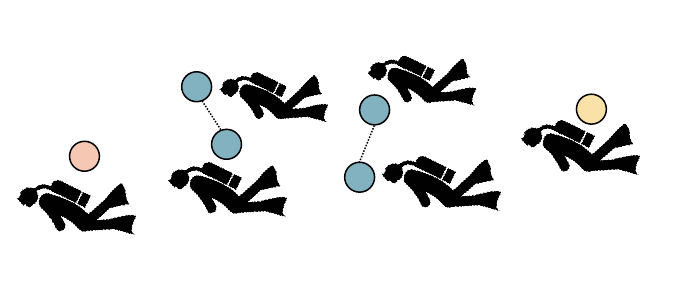}
        \caption{}
    \end{subfigure}

    \bigskip
    \vfill

    \begin{subfigure}{0.47\hsize}
        \centering
        \includegraphics[width=0.8\linewidth]{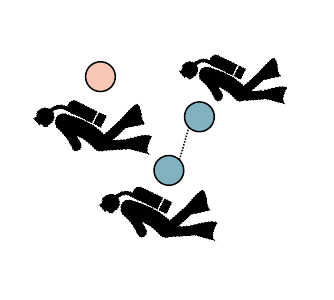}
        \caption{}
    \end{subfigure}%
    \begin{subfigure}{0.47\hsize}
        \centering
        \includegraphics[width=0.8\linewidth]{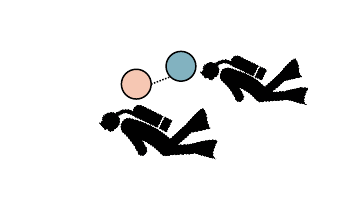}
        \caption{}
    \end{subfigure}

    \bigskip
    \vfill

    \begin{subfigure}{1.0\hsize}
        \centering
        \includegraphics[width=0.8\linewidth]{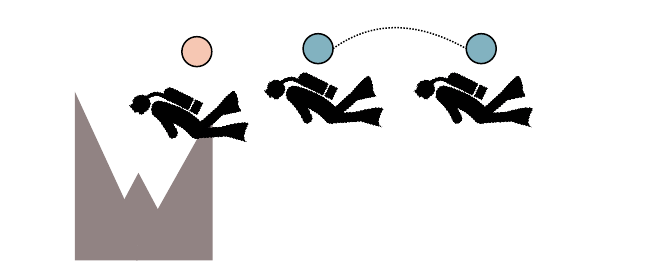}
        \caption{}
    \end{subfigure}

    \caption{Formations of scuba divers.}
    \label{fig:formations}
    \smallskip
    {\small
    Instructor or guide (orange node) in the lead position, followed by pairs of divers (blue nodes; buddy teams indicated by dotted lines), followed by an optional assistant (yellow node) at the rear.
    }
    \vspace{0em plus 0.5em}
\end{figure}

Within a dive group, divers usually hold specific positions \citep{DiveOtterBuddyPositioning, PADIOWDManual}. In guided dives, an instructor or dive guide typically leads the group from the front. The guide often swims at the deepest point of the formation, thereby implicitly limiting the maximum depth. Behind the guide, the participating divers usually move in buddy pairs, maintaining a distance that allows visual contact and quick interaction if necessary. In some cases, an assisting dive master might follow the group in the rear. A typical formation is illustrated in \Cref{fig:formations}(a).

Variations may occur depending on the dive situation, \eg, for groups of three divers (\Cref{fig:formations}(b)) or for an instructor accompanying a single student (\Cref{fig:formations}(c)).
In narrow environments, such as canyons, caves, or shipwrecks, divers may temporarily move in a one-by-one formation (\Cref{fig:formations}(d)).

%% file: parts/background/graphs.tex

\subsection{Graphs}
\label{sec:background_graphs}

Graphs are a key concept for representing structured data, in particular relationships between objects. Objects are represented by vertices, relationships between them by edges \cite[][Chap.~2]{Esser2014}.
For further reading on graphs, see \eg, \citep{Juenger2004}.

A \emph{graph}~$G = (V, E)$ consists of a finite set $V = V(G)$ of \emph{vertices} and a finite set $E = E(G) \subseteq \{\{u,v\} \mid u,v \in V\}$ of \emph{edges}. In an undirected graph, each edge is an \emph{unordered} pair ${u,v}$ connecting two vertices $u,v \in V$.

In the context of scuba diving, each vertex represents a diver and edges represent their relationships, with respect to three dimensions -- physical, communicative, and emergency distance.

In the present framework, undirected graphs are used. For future work, one idea is to introduce directed edges, which are ordered pairs with an orientation, to better represent asymmetric relationships.

This basic graph model can be extended with arbitrary attributes. Vertex attributes may describe properties of divers, such as role or experience level. Edge attributes may describe properties of relationships, in particular by quantifying the three dimensions.
Equivalently, the three dimensions can be represented as separate layers of a multidimensional graph.

From a graph-theoretical perspective, scuba diving graphs are \emph{temporal} networks, \ie, a sequence of graph snapshots $G_t = (V, E_t)$ representing divers and their relationships over time \citep{Doerpinghaus2024, Holme2019}. 

In temporal networks, the set of vertices~$V$ remains stable over time, while only the edges change as relationships evolve. The same holds for scuba diving: During a dive, the group remains the same, and ideally the assigned positions within the group are held. At crowded dive sites, it can happen that two groups cross each other's paths. Divers may become disoriented and accidentally follow the wrong group. From a graph perspective, this is not possible, and likewise in diving practice, this should not happen.

The temporal resolution of such snapshots depends on the available data and the intended analysis. As a rough distinction, snapshots may correspond to broad dive phases, such as descent, exploration, ascent, or safety stop. With sensor-based data, snapshots could be defined at much finer time intervals.

\pagebreak[2]

%% file: parts/related_work.tex

\section{Related Work}
\label{sec:related_work}

To the best of our knowledge, the present work is one of the first attempts to apply CSS methods and graph-based modeling to the analysis of dive groups.

\subsection{Computational Social Science (CSS)}
CSS provides a broad set of methods for analyzing social structures using computational models and network representations. Graph-based approaches have been applied to a wide range of domains, including education, labor markets, historical research, and religious studies -- and now also to scuba diving.

Several studies have explored computational models of small-group interaction.
\cite{Heise2013} introduced the \emph{GroupSimulator}, a computational model based on Affect Control Theory (ACT) that simulates behavior in small social groups.
\cite{Hoey2018} proposed a computational approach to analyze small-group dynamics, extending the \emph{GroupSimulator} by introducing a Bayesian generalization of ACT. Based on this approach, they analyzed interaction during software development on GitHub.

Graph-based models have also been used to analyze the social behavior of animals, modeling movement and leadership patterns as multidimensional interaction networks. \cite{Sampaio2024} published a study on the cooperative hunting behavior of octopuses and fish -- which also form small social groups underwater. This is particularly interesting because octopuses are typically solitary, highly territorial animals. Hunting in multi-species groups that interact collectively is a rare exception.

Social interactions and role distributions in groups are often driven by individual differences. For divers, variations in experience and training strongly influence group dynamics. For animals, such individual differences are less obvious. Multi-species groups provide an opportunity to study social behavior and role differentiation in a more accessible way. In these hunting groups, a clear division of labor emerges: fish explore the environment and patrol the surrounding open water, while the octopus searches caves within the reef. Beyond the physical dimension, current research is exploring the communicative dimension. Observations indicate that octopuses actively indicate when a group should move~\citep{Sampaio2024}.

Beyond these examples, graph-based approaches have been explored in related fields with similar settings -- such as cave rescues or firefighters in smoke-filled buildings. These domains share several characteristics with scuba diving, including small team sizes, spatial coordination requirements, limited visibility, and safety-critical interactions \citep{Desmet2013}.

\subsection{Scuba Diving}

As mentioned above, there has been relatively little research on graph-based approaches in the field of scuba diving.

One context in which graph-like structures are used in diving practice is search and recovery operations. Search patterns (orthogonal or circular grids) are used to locate missing divers, starting from the last known point, systematically covering the area to maximize the probability of finding the missing diver.

In another example, \cite{Shukri2023} proposed an underwater routing framework to guide divers through underwater environments. Besides time, location, and destination, they also took the diver's health status into account as an additional dimension, monitored using smart sensors.


Several data sources could support future empirical analysis of scuba diving interactions.
One important source, especially regarding the emergency dimension, is the annual reports published by the Divers Alert Network (DAN), which provide detailed statistics on diving incidents and fatalities \citep{DANFatalities2021}.

In addition, there are several citizen datasets collecting data from dives. \cite{Brosens2021} published a dataset on dive locations, with a focus on observations of marine life. Online logging platforms\footnote{See, \eg, \emph{\url{https://divelogs.org/}} or \emph{\url{https://scuba.network/}}.} collect logbook entries from divers that include structured metadata (e.g., dive duration, depth, or visibility) as well as optional free-text descriptions of the dive. Visibility information may serve as an indicator for both the physical and communicative distance between divers. Free-text comments may also contain subjective impressions of group dynamics, although such information is often sparsely documented.

An interesting follow-up question is whether divers tend to note negative experiences (\eg, buddy too far away watching a moray) more often than positive ones (\eg, good physical distance). Future research could, provided sufficient data are available, analyze whether there is any bias in this regard.

%% file: parts/graph_representation.tex

\section{Graph Representation}
\label{sec:graph_representation}

In the proposed graph model, each vertex corresponds to a diver and edges capture the relationships between divers along the three dimensions. Accordingly, each pair of vertices is connected by three distinct edges, one for each dimension.

%% file: parts/graph_representation/dimensions.tex

\subsection{Dimensions of Diver Relations}
\label{sec:graph_representation_dimensions}

As mentioned in the introduction, the relationships between divers can be described along three different dimensions: physical distance, communicative distance, and emergency distance.

\subsubsection{Physical Distance}
The most direct relation between divers is their relative spatial position in the water. Physical distance $d_p$ describes how far away, measured in meters, divers are from each other.

If divers move too far apart, assistance becomes more difficult. A typical example occurs when a diver temporarily leaves the group to observe an interesting fish or other underwater object.
Conversely, the physical distance can even be too small, restricting freedom of movement, causing divers to hit one another, or creating the impression of being rushed.

\subsubsection{Communicative Distance}

Communicative distance $d_c$ describes how effectively divers can exchange information during the dive. Underwater communication is typically performed using standardized hand signals, light signals, acoustic signals, or other previously agreed methods.

In a graph representation, communicative distance can be modeled using a quantitative scale, \eg, ranging from negative values (poor communication) to positive values (effective communication).

Communicative and physical distance do not necessarily overlap. Divers can be physically close to each other but still fail to communicate effectively if visibility is poor, \eg, in murky water. Conversely, experienced buddy teams might be able to communicate over greater distances. Therefore, the three dimensions (even though they are strongly correlated) are treated as conceptually distinct.

\subsubsection{Emergency Distance}

The third dimension concerns the ability of divers to assist each other in case of problems. Emergency distance $d_e$ can be measured as the time required for one diver to reach and assist another diver.

Dive training organizations, insurers, and scientific literature commonly distinguish between \emph{minor} and \emph{major} diving mishaps \citep{Ranapurwala2017}.

Minor incidents can usually be managed by the diver or their buddy during a dive, with a low risk of serious injury.
Examples include difficulties equalizing ear pressure, temporary mask flooding, loss of a fin, buoyancy control problems, minor equipment malfunctions, temporary disorientation, or fatigue and muscle cramps.

Major incidents, in contrast, have a high potential to cause severe injury or death; they often require immediate rescue or medical intervention.
The Divers Alert Network (DAN) publishes an annual list of diving fatalities \citep{DANFatalities2021}, with major incidents typically including: loss of gas, contaminated gas, oxygen toxicity, oxygen valve closed, nitrogen narcosis, rapid ascent, environmental hazards, health issues, major equipment failure.

The concept of emergency distance captures how quickly a buddy or another diver can intervene when such situations occur.
In emergency situations, physical distance and reaction time become critical factors. The Professional Association of Diving Instructors (PADI) recommends that a diver should be able to reach their buddy within two seconds~\cite{PADIBuddy}.

%% file: parts/graph_representation/example.tex

\subsection{Example Scenario}
\label{sec:graph_representation_example}

\begin{figure}[tbp]
    \centering
    \includegraphics[
    trim=0 1.0cm 1.75cm 0, 
    clip,
    width=0.55\linewidth
    ]{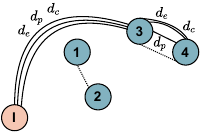}
    \caption{Graph representation of an example scenario.}
    \label{fig:example}
    \smallskip
    {\small
    Group of four divers (blue nodes; buddy teams indicated by dotted lines) following an instructor (I, orange node). Edge labels denote the physical distance $d_p$, communicative distance $d_c$, and emergency distance $d_e$ between two divers.
    }
    \vspace{0em plus 0.5em}
\end{figure}

A realistic and frequently occurring situation during a dive can illustrate the three dimensions. Consider a group of four divers following an instructor, as illustrated in \Cref{fig:example}.

Suppose that Diver~3 indicates a problem with ear equalization (standard hand signal: wiggling one hand, \enquote{I have a problem}, then pointing to the ears). To resolve the issue, Diver~3 slightly ascends.

Diver~4, acting as a responsible buddy, follows this movement and maintains a small physical distance $d_p$. By staying close, Diver~4 ensures effective communication $d_c$ (\eg, by signaling \enquote{Take your time.}, flat hand, and \enquote{Are you okay?}, thumb and index finger forming an O). The reduced physical distance also minimizes the emergency distance $d_e$, allowing immediate intervention if the situation worsens.

The instructor, positioned at the front of the group and typically a bit deeper, faces a more complex decision. While maintaining responsibility for the entire group, the instructor must quickly assess the severity of the situation and whether Diver~3 can resolve the issue independently.
In practice, the instructor would gradually reduce the distances along all three dimensions: Slowing down or temporarily stopping the group to decrease physical distance, monitoring and possibly communicating with the diver to reduce communicative distance, and positioning oneself to be able to intervene rapidly in case of a major emergency. This example illustrates how the three dimensions interact dynamically.

%% file: parts/graph_representation/vertical_limitations.tex

\subsection{Vertical Limitations}
\label{sec:graph_representation_limitations}

As discussed in \Cref{sec:background_graphs}, scuba diving graphs are \emph{temporal} networks, where each graph corresponds to a snapshot at a specific point in time. However, consecutive snapshots cannot differ arbitrarily, as vertical movement is constrained by physical and regulatory limits.

Key limitations include the maximum depth, which defines how deep a diver may descend.
In addition, divers are required to perform a safety stop at approximately five meters before surfacing to reduce the risk of decompression sickness (DCS).
Ascent rates are strictly limited~\cite{CMASAscentRate}, constraining how quickly a diver can change depth over time. Modern dive computers can measure the ascent rate in real time and warn a diver by beeping in case this constraint is violated.

\begin{figure}[tbp]
    \centering
    \begin{subfigure}[t]{0.6\linewidth}
        \centering
        \includegraphics[width=\linewidth]{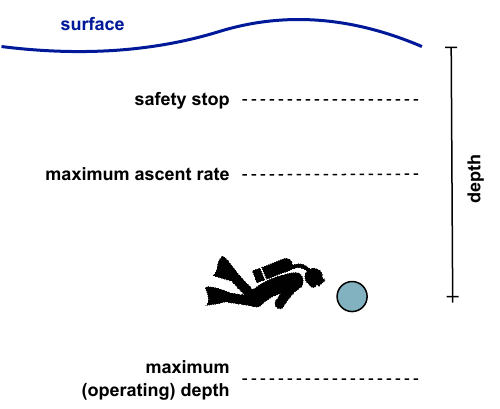}
        \caption{}
    \end{subfigure}
    ~\\
    \vspace{2em plus 1em}
    ~\\
    \begin{subfigure}[t]{\linewidth}
        \centering
        \includegraphics[width=\linewidth]{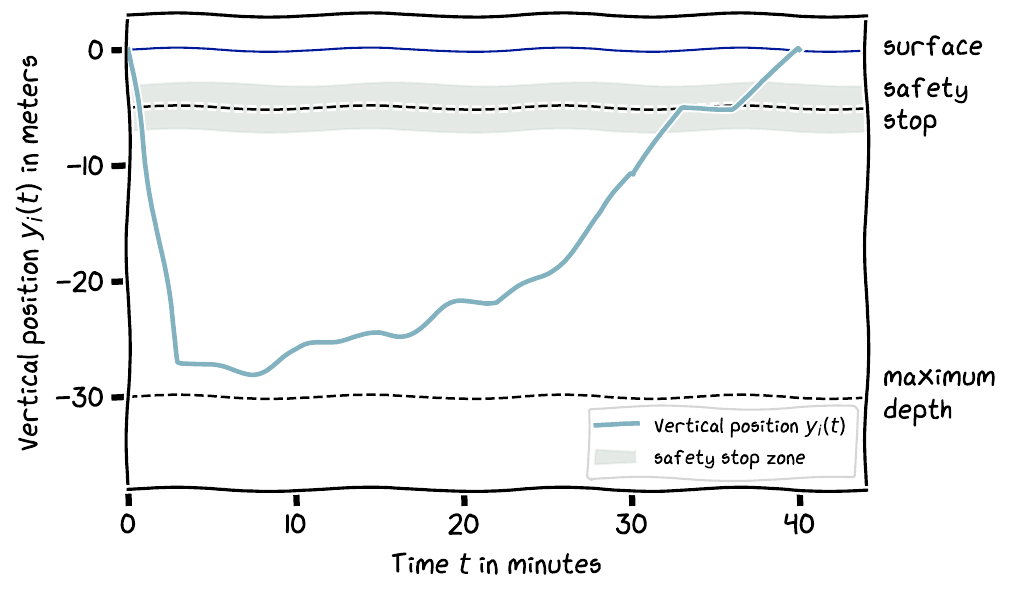}
        \caption{}
    \end{subfigure}

    \caption{Vertical limitations in scuba diving.}
    \label{fig:vertical}
    \smallskip
    {\small
        (a):~Single temporal snapshot with vertical constraints.\\
        (b):~Dive profile showing the depth over time.
    }
    \vspace{0em plus 0.5em}
\end{figure}

\Cref{fig:vertical} illustrates these vertical limitations. \Cref{fig:vertical}(a) shows a single temporal snapshot and the existing vertical constraints. Let $G_t$ denote the current snapshot, and let $y_i(t)$ denote the depth of diver $i$ at time $t$, measured negatively downward. Then, the diver's depth $y_i(t+1)$ in the next temporal snapshot $G_{t+1}$ must satisfy the constraints described above. This results in a dive profile with a specific depth over time, as illustrated in \Cref{fig:vertical}(b).

These constraints limit how rapidly the spatial positioning of the dive group -- and thus, the graph -- can change between consecutive time steps.
If a diver fails to observe the ascent rate, their dive computer might block them for 24 hours (a very bad scenario for many divers) or the diver might develop symptoms of decompression sickness (even worse).
Therefore, vertical limitations are important boundaries on the temporal dynamics of scuba diving graphs.

%% file: parts/conclusion.tex

\section{Conclusion}
\label{sec:conclusion}

\glsreset{CSS}
This paper presented a conceptual approach for representing groups of scuba divers as graphs. The proposed model focuses on three central dimensions of diver relations: physical distance, communicative distance, and emergency distance. In this way, the paper demonstrates how methods from \gls{CSS} can be transferred to an unusual but structurally suitable domain such as scuba diving.

\pagebreak[2]

The proposed representation is currently theoretical and intended as a first step toward a more systematic computational description of diver interactions. Its main contribution is to show that scuba diving groups can be understood as small, structured social networks whose relations are shaped not only by spatial position but also by communication and emergency responsiveness, with these safety-critical factors having immediate practical relevance. This perspective may help to connect practical diving behavior with established concepts from graph theory and social network analysis.

A particularly important question for future analyses is how deviations from the intended physical formation affect the other two dimensions. Increased physical distance may reduce communication quality and slow down emergency response. In special cases where a temporary deviation from the usual formation is necessary, \eg, when passing through a narrow canyon, it is interesting to analyze how to preserve the smallest possible communicative and emergency distance.

\subsection{Limitations and Future Work}
\label{sec:limitations_future_work}

The proposed model is conceptual and has not yet been validated empirically.
A further limitation is that the communicative and emergency distances are difficult to measure directly in practice and partly depend on subjective assessments. They can only be measured indirectly via proxies, such as visibility, line of sight, communication success, or response time.

One important step for future work is the collection of suitable data. Many divers already document their dives in digital logbooks, including standard metadata such as duration, maximum depth, and visibility, and sometimes also free-text comments. Some logging applications additionally ask divers to rate selected criteria on simple scales. Such tools could be extended to include structured debriefing questions about position keeping, communication, buddy awareness, and incidents during the dive. These evaluations could provide practical data for the three proposed dimensions.

The availability of data would generally help to populate the graph with additional information and thereby provide a more realistic representation of the dive. This would be an important step toward recording dives over time in a structured form for subsequent analysis.

The approach may also benefit from advances in dive computer technology. Modern dive computers already record detailed information such as depth profiles and, in some cases, physiological metrics such as heart rate (which may serve as an indicator of emergency situations). If data from two or more divers could be synchronized and jointly analyzed, this would allow a more precise reconstruction of relative positions and group dynamics during a dive.

Future work may also extend the model by introducing additional dimensions. Examples include air consumption (as a key factor in diving) or experience level (often linked to a specific responsibility status within the group).

Directed graphs could be useful for representing asymmetric relations, such as instructor--student or experienced diver--novice diver constellations.

The present work is purely \emph{descriptive}. It focuses on characterizing the social behavior of divers with respect to the three proposed dimensions.
A possible future step would be the development of a \emph{predictive} approach that supports decision-making in specific situations. For example, such a model could help to estimate when an instructor should temporarily leave the lead position in order to assist a diver showing early signs of difficulty (see \Cref{sec:graph_representation_example}).
\emph{Simulations} could be used to explore how changes in group positioning, for example when passing through a narrow canyon (see \Cref{fig:formations}(d)), affect the three dimensions in order to further optimize the positioning.

An advanced concept could also take into account the \emph{reasons} for specific positioning. In a debriefing, divers could reflect on the phases of the dive when positioning was suboptimal, considering the three dimensions. They could identify potential contributing factors, such as current, difficulties with buoyancy control, or physical reasons.

While such developments are beyond the scope of this paper, they illustrate the broader potential of graph-based computational models for improving both the understanding and the safety of scuba diving.